\documentclass[aps, prl, superscriptaddress, preprintnumbers, twocolumn, nofootinbib]{revtex4}

\usepackage{amsmath}	
\usepackage{amsthm}	
\usepackage{amssymb}	
\usepackage{datetime}
\usepackage{graphicx}
\usepackage{color}
\usepackage{verbatim}
\usepackage{multirow}

\usepackage[bookmarks, breaklinks, plainpages=false, pdfpagelabels, colorlinks,linkcolor=darkblue,citecolor=darkblue,filecolor=darkblue,urlcolor=darkblue]{hyperref} 

\definecolor{grey}{rgb}{0.4,0.4,0.4}
\definecolor{dullmagenta}{rgb}{0.4,0,0.4}
\definecolor{darkblue}{rgb}{0,0,0.4}
\definecolor{midblue}{rgb}{0,0,0.5}
\definecolor{midred}{rgb}{0.5,0,0}
\definecolor{orange}{rgb}{1,0.5,0}
\definecolor{lightbrown}{rgb}{0.75,0.5,0.25}
\definecolor{tan}{cmyk}{0.14,0.42,0.56,0}
\definecolor{djunglegreen}{cmyk}{0.99,0,0.52,0}
\definecolor{lightgreen}{rgb}{0,1,0}
\definecolor{olivegreen}{cmyk}{0.64,0,0.95,0.40}
\definecolor{midgreen}{rgb}{0.0,0.675,0.0}
\definecolor{darkgreen}{rgb}{0,0.5,0}

\renewcommand{\.}{\hspace{0.5mm}}

\newcommand{\ra}{\ensuremath{\rightarrow}}

\newcommand{\grm}{\ensuremath{\mathrm{g}}}

\newcommand{\srm}{\ensuremath{\mathrm{s}}}

\newcommand{\Ocal}{\ensuremath{\mathcal{O}}}

\newcommand{\cm}{\ensuremath{\rm cm}}

\newcommand{\GeV}{\ensuremath{\mathrm{GeV}}}

\newcommand{\eg}{e.g.}
\newcommand{\ie}{i.e.}

\newcommand{\cf}{cf.} 

\let\baraccent=\= 
\renewcommand{\=}[1]{\stackrel{#1}{=}} 

\theoremstyle{definition}

\theoremstyle{remark}

\settimeformat{ampmtime}

\usepackage{float}

\begin{document}

\title{Decaying Dark Matter in Halos of Primordial Black Holes}

\author{Florian K{\"u}hnel}
\email{florian.kuhnel@fysik.su.se}
\affiliation{Department of Physics,
	School of Engineering Sciences,\\
	KTH Royal Institute of Technology,
	AlbaNova University Center,\\
	Roslagstullsbacken 21,
	SE--106\.91 Stockholm,
	Sweden}
\affiliation{The Oskar Klein Centre for Cosmoparticle Physics,
	AlbaNova University Center,\\
	Roslagstullsbacken 21,
	SE--106\.91 Stockholm,
	Sweden}

\author{Tommy Ohlsson}
\email{tohlsson@kth.se}
\affiliation{Department of Physics,
	School of Engineering Sciences,\\
	KTH Royal Institute of Technology,
	AlbaNova University Center,\\
	Roslagstullsbacken 21,
	SE--106\.91 Stockholm,
	Sweden}
\affiliation{The Oskar Klein Centre for Cosmoparticle Physics,
	AlbaNova University Center,\\
	Roslagstullsbacken 21,
	SE--106\.91 Stockholm,
	Sweden}
\affiliation{University of Iceland, 
	Science Institute, 
	Dunhaga 3, 
	IS--107 Reykjavik, 
	Iceland}

\date{\formatdate{\day}{\month}{\year}}

\begin{abstract}
We investigate photon signatures of general decaying dark-matter particles in halos of primordial black holes. We derive the halo-profile density and the total decay rate for these combined dark-matter scenarios. For the case of axion-like particles of masses below $\Ocal( 1 )\.{\rm keV}$, we find strong bounds on the decay constant which are several orders of magnitude stronger than the strongest existing bounds, for all halo masses above $\Ocal( 10^{-5} )$ solar masses. Using future X-ray measurements, it will be possible to push these bounds on such combined dark-matter scenarios even further.
\end{abstract}

\maketitle

{\it Introduction\,---\,}In the standard model of cosmology, the energy density of the Universe consists of approximately $25\,$\% in the form of a pressureless, nearly perfect fluid of non-relativistic objects, so-called (cold) dark matter. A large number of potential dark-matter candidates have been proposed so far. The perhaps most well-studied class is constituted by hypothetical particles which only weakly interact with the other standard-model particles. Amongst this variety, there are so-called WIMPs (\cf~Ref.~\cite{Steigman:1984ac}), sterile neutrinos (see Ref.~\cite{Dodelson:1993je} for an early discussion on their role as dark-matter components and Refs.~\cite{Boyarsky:2009ix, Adhikari:2016bei, Abazajian:2017tcc, Bernal:2017kxu}), axions \cite{Peccei:1977ur, Peccei:1977hh, Weinberg:1977ma, Wilczek:1977pj}, and axion-like particles (ALPs) \cite{Chikashige:1980ui}. The latter, whose characteristics we will use in this work, constitutes a class of pseudo Nambu-Goldstone bosons which are coupled to photons. For axions, their mass and decay constant are related, while this is generally not the case for ALPs.

Besides microscopic candidates like ALPs, dark matter might also be constituted by macroscopic objects such as primordial black holes (PBHs) \cite{1967SvA....10..602Z, Carr:1974nx}.\footnote{There exist yet two other possibilities for macroscopic dark matter, namely nuclear-density objects (\cf~Refs.~\cite{Witten:1984rs, Lynn:1989xb}), and as so-called ultracompact mini-halos (UCMHs) \cite{Ricotti:2009bs}.} These are black holes which have been produced in the very early Universe. The interest in PBH constituting parts of the dark matter \cite{1975Natur.253..251C} has been revived recently \cite{Carr:2009jm, Bird:2016dcv, Carr:2016drx, Clesse:2016vqa, Green:2016xgy, Kuhnel:2017pwq, Carr:2017jsz, Carr:2017edp, Carr:2018poi, Kuhnel:2019xes, Carr:2019kxo}, in particular through the gravitational-wave discovery of black-hole binary mergers \cite{Abbott:2016blz, Abbott:2016nmj}. The possible PBH formation mechanisms are very diverse and there is a large number of scenarios, which lead to their formation. All of these have in common that they require some mechanism to generate large overdensities.

Even though most of the emphasis in dark-matter research has been focused on one-component scenarios, models with more than one component have been investigated, including mixed types of both microscopic as well as macroscopic nature. On the one hand, a small fraction of PBHs could provide seeds for super-massive black holes in the galactic centres \cite{Bean:2002kx}. On the other hand, in view of the fact that it appears difficult, although not impossible, to have the entire dark matter in the form of PBHs or UCMHs (\cf~Ref.~\cite{Carr:2016drx} including a summary of relevant constraints), the class of particle dark matter provides a vital supplementary and major candidate.

In all of those combined scenarios, the particles will be gravitationally bound to the PBHs. This could lead to strong decay \cite{Kuhnel:2017ofn} and/or annihilation signatures \cite{Eroshenko:2016yve, Boucenna:2017ghj}.\\[-2mm]

\begin{figure}
	\begin{center}
	\includegraphics[width = \linewidth]{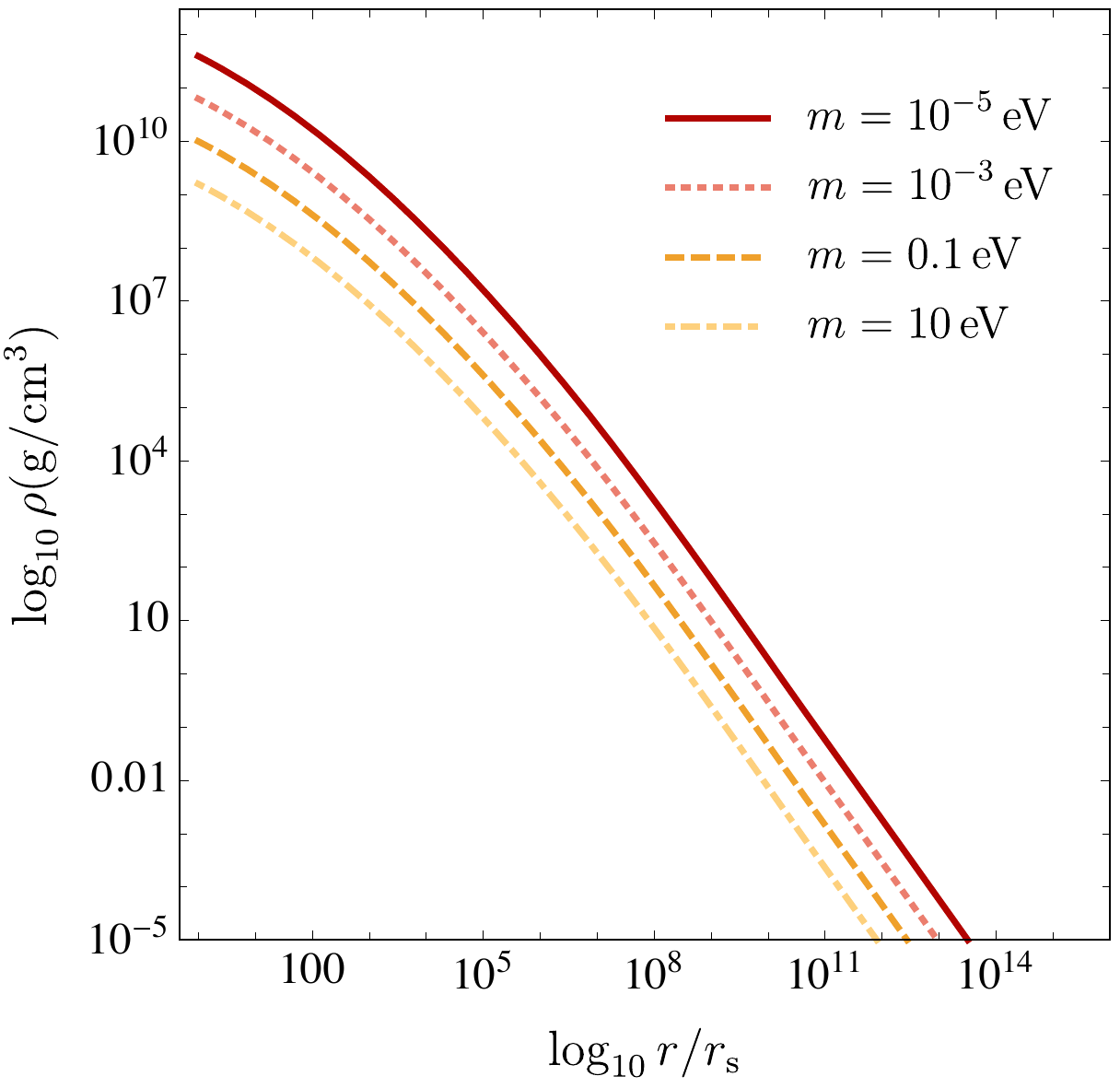}
	\caption{
			Halo-profile density $\rho$ (in units of $\rm g / cm^{3}$)
			around a PBH as a function of $r / r_{\rm s}$, 
			assuming the formation mechanism discussed 
			in Refs.~\cite{Eroshenko:2016yve, Boucenna:2017ghj}.
			The values of the particle mass $m$ for each curve are
			$m = 10^{-5}\.{\rm eV}$ (maroon solid curve), 
			$m = 10^{-3}\.{\rm eV}$ (pink dotted curve), 
			$m = 0.1\.{\rm eV}$ (orange dashed curve), 
			and
			$m = 10\.{\rm eV}$ (yellow dot-dashed curve),
			respectively.
			}
	\label{fig:Profil}
	\end{center}
\end{figure}

{\it Halos\,---\,}As mentioned above, in a combined dark-matter scenario consisting of a large fraction of particles and a small fraction of PBHs, the former will be gravitationally bound to the latter. For WIMPs, this has been studied by Eroshenko \cite{Eroshenko:2016yve} and the authors \cite{Boucenna:2017ghj}. However, this formation mechanism, which happens in the radiation-dominated epoch, is not specific to any particular WIMPs model. In fact, the essential ingredients are the mass and the velocity distribution of the particles. Hence, we will generalize the results to investigate general halo formation and follow Ref.~\cite{Boucenna:2017ghj}, wherein the technical details can be found. Figure~\ref{fig:Profil} presents the halo-profile density as a function of radius $r$ (in units of the Schwarzschild radius $r_{\rm s} \simeq 2.95 \cdot 10^{3}\.{\rm km}$) for accreted particles of mass $m \in \{ 10^{-5},\,10^{-3},\,0.1,\,10 \} \.{\rm eV}$ around a PBH of a solar mass $M_{\odot}$, assuming a Maxwellian velocity profile. Note that the calculation leading to Fig.~\ref{fig:Profil} only relies on gravitational dynamics, implying that the halo-profile density solely depends on the particle mass $m$. In the case the particle dark matter is constituted by a number of different species (with different masses), the halo-profile density changes in a non-trivial way. In the following, we will assume that each halo is entirely constituted by a single particle type.

As can be observed from Fig.~\ref{fig:Profil}, lighter particles lead to a more extended halo. Outside of the halo's core, its profile follows $\rho( r ) \propto m^{- 2 / 5}\,r^{- 3 / 2}$, which can be estimated from Fig.~\ref{fig:Profil}.\footnote{It must be noted that there is a debate (see Ref.~\cite{Adamek:2019gns}) about the specific form of the halo profile. Our calculation is based on previous results of Ref.~\cite{Boucenna:2017ghj}. For decays, the specific form does not matter; it is merely the total number of decaying particles (determined by the halo mass) which affects the decay rate.} The behaviour of the halo-profile density as a function of $m$ derives from the fact that lighter particles become non-relativistic at a later time than heavier ones, and hence the background density $\rho$ will be lower. The radius of gravitational influence $r_{\rm infl}$ of the PBHs scales as $r_{\rm infl} \sim \rho^{- 3 / 2}$ (see Ref.~\cite{Boucenna:2017ghj}). This leads to larger structures for smaller particle masses. At matter-radiation equality, the mass $M$ of the halo is comparable to that of the PBH, see the discussion in Ref.~\cite{Adamek:2019gns}. After matter-radiation equality, the growth of the halo is roughly linear in redshift, leading to approximately a factor of 1000 difference in mass. Thus, essentially all the halo mass is constituted by the sum of the particle masses, meaning that the number $N$ of particles within the halo is roughly given by $N = M / m$. We will mainly be interested in the case in which the dark matter is essentially constituted by ALPs, \ie~when $f_{\rm ALP} \equiv \rho_{\rm ALP} / \rho_{\rm DM} \approx 1 \gg f_{\rm PBH} \equiv \rho_{\rm PBH} / \rho_{\rm DM}$, where $\rho_{\rm ALP}$, $\rho_{\rm PBH}$, and $\rho_{\rm DM}$ are the energy densities of ALPs, PBHs, and dark matter, respectively. Furthermore, we will assume that all of the ALPs are bound to the PBHs in the halos, \ie~$f_{\rm halo} = f_{\rm ALP} + f_{\rm PBH} \approx f_{\rm ALP}$.\\[-2mm]

\begin{figure}
\begin{center}
	\includegraphics[width = \linewidth]{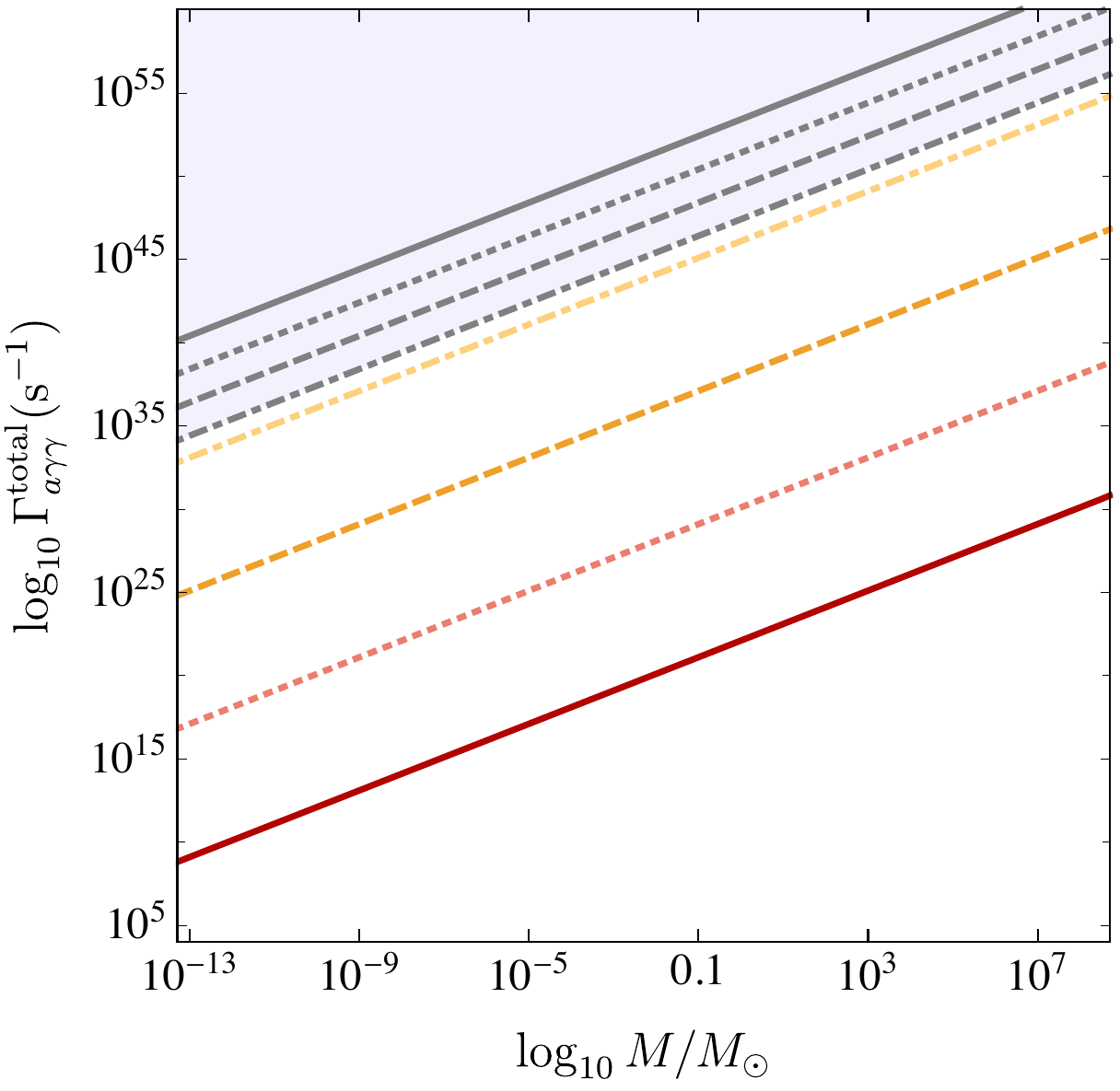}
	\caption{Total decay rate $\Gamma_{a\gamma\gamma}^{{\rm total}}$ 
			as a function of halo mass $M$
			in units of the solar mass $M_{\odot}$
			for different values of the particle mass $m$ (see Fig.~\ref{fig:Profil}).
			The grey-shading indicates the corresponding regions in which 
			the halos would have already decayed by now. The different black lines
			correspond to the colored lines with the same line style.
			}
	\label{fig:PlotGamma}
	\end{center}
\end{figure}

{\it Decay\,---\,}For decay signatures, distinct from annihilations, and unless the halo is extremely close to the telescope, it is practically point-like, and hence its total mass matters rather than its concrete density profile. Given an individual decay rate, the total decay rate\footnote{In the case of $f_{\rm ALP} < 1$, the total decay rate is replaced by $\Gamma^{\rm total} \ra f_{\rm ALP}\.\Gamma^{\rm total}$, and therefore lowered, provided the remainder of the halo particles does neither decay nor annihilate. Otherwise, there will be additional contributions.} is readily obtained using $\Gamma^{\rm total} = N\,\Gamma$. For ALPs (see Sec.~111 of Ref.~\cite{Tanabashi:2018oca} for a recent review), we may write
\begin{align}
	\Gamma
		\equiv
	\Gamma_{a\gamma\gamma}
		&=
					\frac{ G_{a\gamma\gamma}^{2} }{ 64\.\pi }\,m^{3}
					\; ,
					\label{eq:Gamma}
\end{align}
where $G_{a\gamma\gamma}$ is the decay constant. For the QCD axion, Eq.~\eqref{eq:Gamma} simply becomes
\begin{align}
	\Gamma_{a\gamma\gamma}
		&\simeq
					1.1 \cdot 10^{-24}\,
					\left(
						\frac{ m }{ {\rm 1\,eV} }
					\right)^{\!5}\;\srm^{-1}
		\propto
					m^{5}
					\; .
					\label{eq:GammaQCDaxion}
\end{align}
For sterile neutrinos, a similar expression holds (see Ref.~\cite{Kuhnel:2017ofn}). Figure~\ref{fig:PlotGamma} shows the total decay rate $\Gamma_{a\gamma\gamma}^{{\rm total}}$ as a function of the halo mass $M$ for different values of the particle mass $m$, where we assumed for illustrational purpose $G_{a\gamma\gamma} = G^{\rm QCD}_{a\gamma\gamma}$. Furthermore, we have that $\Gamma_{a\gamma\gamma}^{{\rm total}} \propto M\,m^{4}$, which holds if $M$ is fixed. However, if $N$ is fixed, we instead have $\Gamma_{a\gamma\gamma}^{{\rm total}} \propto N\,m^{5}$. Explicitly for the total decay rate, we obtain
\begin{align}
	\Gamma_{a\gamma\gamma}^{{\rm total}}
		&\simeq 1.3
					\cdot 10^{42}\,
					\left(\frac{M}{M_{\odot}}\right)
					\left(\frac{m}{{\rm 1\,eV}}\right)^{\!4}\;\srm^{-1}
					\; ,
\end{align}
which can easily be computed for, \eg
\begin{align}
\begin{split}
	M
		&\in \{ 10^{-5},\;0.1,\;10^{3},\;10^{8} \} \.M_{\odot}
					\; ,
					\\[1mm]
	m
		&\in \{ 10^{-5},\,10^{-3},\,0.1,\,10 \} \.{\rm eV}
					\; ,
\end{split}
\end{align}
that indeed agrees with the results of Fig.~\ref{fig:PlotGamma}.\\[-2mm]

{\it Constraints\,---\,}In Ref.~\cite{Kuhnel:2017ofn}, we proposed and investigated a scenario in which the dark matter is constituted by halos of sterile neutrinos around PBHs. Therein, we studied the possibility that in a certain observational time frame with a certain probability one of those compact objects propagates at a given minimum distance near the telescope. Through the halo's nearness, its radiation may dominate the photon flux from other sources onto the telescope. If the halo's minimum distance is small enough, its signature will be detected.

It is now tempting to generalize this set-up. Applied to ALPs, we derive new limits on the maximally-allowed decay constant $G_{a\gamma\gamma}$. Reference \cite{Arias:2012az} provides a variety of bounds for an extended mass range. We compare the observed photon fluxes to those originating from the decays of the ALP halos using the methodology of Ref.~\cite{Kuhnel:2017ofn}, described in the previous paragraph. Utilizing the relation from the total decay rate to the decay constant [see Eq.~\eqref{eq:Gamma}], we obtain constraints for the latter.

Concretely, for a selection of halos with different halo masses ($M \in \{ 10^{-5},\;0.1,\;10^{3} \} \.M_{\odot}$), we compare their photon fluxes to the one received by the telescope assuming no halos. Furthermore, we suppose that the local dark-matter density is spatially homogeneous and takes a value of $0.3\.{\rm GeV}\.{\rm cm}^{-3}$. This determines the average distance $d$ between two halos:
\begin{align}
	d
		&\approx
					1.2 \cdot 10^{8}
					\left( \frac{ M }{ 1\.\grm } \right)^{\!1 / 3}
					\, {\cm}
					\, .
					\label{eq:d-average}
\end{align}
The velocity distribution of the halos is assumed to be Maxwellian. As described above, subject to this distribution with a certain probability $P$, a halo will move near the telescope and shed photons onto it. As we showed in Ref.~\cite{Kuhnel:2017ofn}, $P$ is approximately given by
\begin{align}
	P
		&\simeq
					\frac{ 16 }{ \pi^{2} }
					\left(
						\frac{ \theta\,\phi }{ 2\.\pi^{2} }
					\right)\mspace{1mu}
					\arcsin^{2}\mspace{-3.5mu}
					\left(
						\frac{ 2\.r_{\Phi} }{ d }
					\right)
					\, ,
					\label{eq:P}
\end{align}
where $\theta \in [0,\.\pi]$ and $\phi \in [0,\.2\pi)$ are the opening angles of a detector. Above, $r_{\Phi}$ is the distance from the detector such that a certain flux $\Phi^{}_{\!A}$ through its effective area $A$ is observed; it is given by \cite{Kuhnel:2017ofn}
\begin{align}
	r_{\Phi}
		&\simeq
					\sqrt{	
						\frac{ A }{ 2\.\pi\.\Phi^{}_{\!A} }\;
						\Gamma_{a\gamma\gamma}^{{\rm total}}
					}
					\; .
					\label{eq:rPhi}
\end{align}
For a given observational time, it is then easy to determine the photon flux of the halos and compare it to that of the background.\footnote{Note that, as opposed to the case of sterile-neutrino decay \cite{Kuhnel:2017ofn}, in which only one photon is emitted in each process, the photon isotropic flux is twice as large.} By virtue of Eq.~\eqref{eq:Gamma}, this can thus be used to constrain $G_{a\gamma\gamma}$.

\begin{figure}
\begin{center}
	\includegraphics[width = \linewidth]{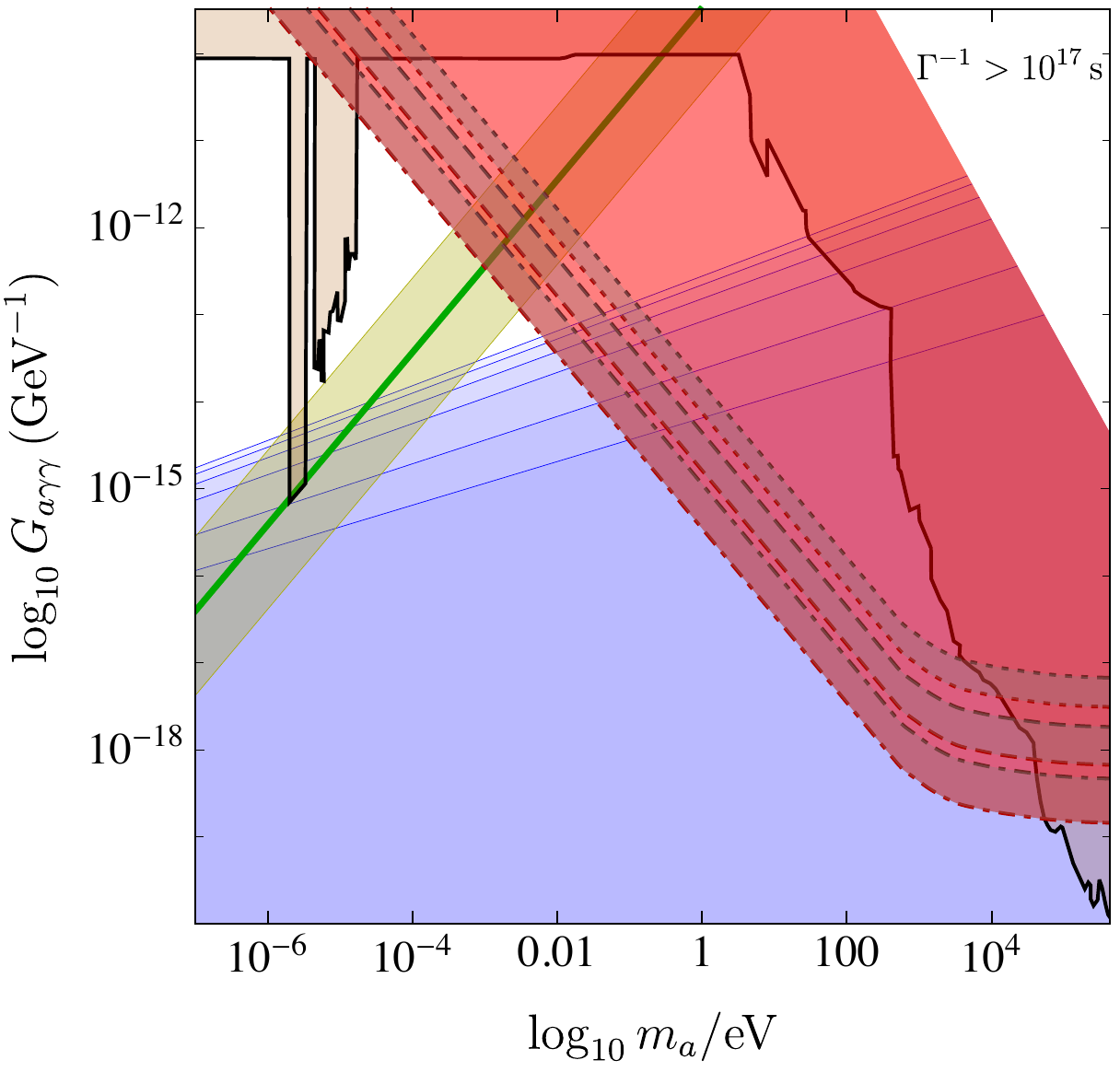}
	\caption{Combined bounds on the decay constant $G_{a\gamma\gamma}$ 
			(in units of $\rm GeV^{-1}$) 
			as a function of ALP mass $m_a$ (in units of $\rm eV)$.
			Note that the corresponding photon energy is half the ALP mass.
			The grey-shaded area (surrounded the black solid curve) 
			indicates bounds for the standard scenario without halos,
			the light-green-shaded band shows various axion models around
			the QCD-axion scenario (green line), and 
			the blue-shaded regions indicate different thermal-mass models 
			using the realignment mechanism
			(see Ref.~\cite{Arias:2012az} for details).
			The red-shaded regions represent our results for three different 
			choices of halo masses $M$ in the combined dark-matter scenario, 
			which are
			$M = 10^{-5}\.M_{\odot}$ (upper band), 
			$M = 0.1 \.M_{\odot}$ (middle band), 
			and
			$M = 10^{3}\.M_{\odot}$ (lower band), respectively.
			In each band, the lower (more restrictive) boundary 
			utilizes a halo dark-matter fraction of one, whereas 
			the upper (less restrictive) boundary utilizes the maximally allowed
			dark-matter fraction from various observations (see main text for details).
			The white area in the upper-right corner depicts the region	
			of parameter space in which the ALP halos have already decayed.\\
			}
	\label{fig:X-ray-Bounds}
	\end{center}
\end{figure}

As a function of the ALP mass $m_{a}$, our results are depicted in Fig.~\ref{fig:X-ray-Bounds}. In this figure, it can be observed that an increase of $M$ leads to a decrease of the constraint line of $G_{a\gamma\gamma}$. Also, smaller values of the energy lead to stronger constraints. The physical reason for this is that, for a fixed $M$, the number $N_{a} = M / m_{a}$ of ALPs within the halo is increasing with decreasing $m_{a}$, whereas the background ray flux is given and fixed, the single halo moving in the vicinity of the telescope contains more decaying particles the smaller their mass is. This extra factor of $1 / m_{a}$ is responsible for the increased detection prospects towards smaller mass. 

In Fig.~\ref{fig:X-ray-Bounds}, we observe that for $m_{a}$ below $\Ocal( 10 )\.{\rm keV}$, we obtain bounds on $G_{a\gamma\gamma}$ which are several orders of magnitude stronger than the strongest existing bounds, for all $M$ above $\Ocal( 10^{-5} )$ solar masses if one assumes a halo dark-matter fraction of one. In particular, for $1\.{\rm keV} \leq m_{a} \leq 300\.{\rm keV}$, we find that the following values of $G_{a\gamma\gamma}$ will, at least, be excluded
\begin{align}
\begin{split}
	G_{a\gamma\gamma}
		&\gtrsim
					10^{-17.0}\, \GeV^{-1} \quad \mbox{for} \quad M \geq 10^{-5} \. M_{\odot}
					\, ,
					\\[1mm]
	G_{a\gamma\gamma}
		&\gtrsim
					10^{-17.7}\, \GeV^{-1} \quad \mbox{for} \quad M \geq 0.1 \. M_{\odot}
					\, ,
					\label{eq:Excluded-Values-of-Gagammagamma}
					\\[1mm]
	G_{a\gamma\gamma}
		&\gtrsim
					10^{-18.3}\, \GeV^{-1} \quad \mbox{for} \quad M \geq 10^{3} \. M_{\odot}
					\, .
\end{split}
\end{align}
These bounds are slightly relaxed if a lower halo dark-matter fraction is accounted for. For the three considered masses, the constraints come from the EROS/OGLE microlensing survey \cite{Allsman:2000kg, Tisserand:2006zx, Wyrzykowski:2011} (for $M \in \{ 10^{-5}$, $0.1 \}\.M_{\odot}$) and X-ray emission from accretion gas around PBHs \cite{Inoue:2017csr} (for $M = 10^{3}\.M_{\odot}$).\footnote{It must be noted that these constraints are subject to assumptions whose validity is hard to quantify.} Taken at face value, these yield the maximally allowed halo dark-matter fractions: $f_{\rm halo}( M = 10^{-5}\.M_{\odot} ) \approx 0.1$, $f_{\rm halo}( M = 0.1 \.M_{\odot} ) \approx 0.05$, and $f_{\rm halo}( M = 10^{3}\.M_{\odot} ) \approx 0.03$. Approximately, the constraint curves in Fig.~\ref{fig:X-ray-Bounds} scale as $f_{\rm halo}^{- 1 / 3}$.

The obtained bounds in this paper can be regarded a conservative, as we have not included ALP-to-photon conversion due to magnetic fields. The latter can occur in two distinct situations:~{\it internally}, for instance for charged and rotating black holes, and {\it externally}, for instance from galactic magnetic fields. Either case only increases the photo emission from the halo objects, and hence strengthens the bounds. It would be interesting to investigate these instances in the future, but it is beyond the scope of the present paper.\\[-2mm]

{\it Conclusions\,---\,}We have investigated decay signatures from a two-component dark-matter scenario in which most of the dark matter is constituted by axion-like particles (ALPs) complemented by primordial black holes (PBHs). We have studied how the former accrete around the latter and calculated the halo profile (shown in Fig.~\ref{fig:Profil}). Then, we have studied the decay signatures (visualised in Fig.~\ref{fig:PlotGamma}) from which we have derived bounds on the decay constant (depicted in Fig.~\ref{fig:X-ray-Bounds}). We have found that this combined scenario leads to detection prospects which, for small ALP masses less than or equal to $\Ocal( 1 )\,{\rm keV}$ and for halos heavier than $10^{-5}\.M_{\odot}$, are far better than the pure ALP scenario.\\

{\it Acknowledgement\,---\,}{We thank the anonymous referee for valuable remarks. F.K.~acknowledges~support~by the Swedish Research Council (Vetenskapsr{\aa}det) through contract No.~638-2013-8993 and the Oskar Klein Centre for Cosmoparticle Physics and T.O.~acknowledges support by Swedish Research Council (Vetenskapsr{\aa}det) through contract No.~2017-0393 and the KTH Royal Institute of Technology for a sabbatical period at the University of Iceland.



\begin{thebibliography}{39}
\expandafter\ifx\csname natexlab\endcsname\relax\def\natexlab#1{#1}\fi
\expandafter\ifx\csname bibnamefont\endcsname\relax
   \def\bibnamefont#1{#1}\fi
\expandafter\ifx\csname bibfnamefont\endcsname\relax
   \def\bibfnamefont#1{#1}\fi
\expandafter\ifx\csname citenamefont\endcsname\relax
   \def\citenamefont#1{#1}\fi
\expandafter\ifx\csname url\endcsname\relax
   \def\url#1{\texttt{#1}}\fi
\expandafter\ifx\csname urlprefix\endcsname\relax\def\urlprefix{URL }\fi
\providecommand{\bibinfo}[2]{#2}
\providecommand{\eprint}[2][]{\url{#2}}

\bibitem[{\citenamefont{Steigman and Turner}(1985)}]{Steigman:1984ac}
\bibinfo{author}{\bibfnamefont{G.}~\bibnamefont{Steigman}} \bibnamefont{and}
   \bibinfo{author}{\bibfnamefont{M.~S.} \bibnamefont{Turner}},
   \bibinfo{journal}{Nucl. Phys. B} \textbf{\bibinfo{volume}{253}},
   \bibinfo{pages}{375} (\bibinfo{year}{1985}).

\bibitem[{\citenamefont{Dodelson and Widrow}(1994)}]{Dodelson:1993je}
\bibinfo{author}{\bibfnamefont{S.}~\bibnamefont{Dodelson}} \bibnamefont{and}
   \bibinfo{author}{\bibfnamefont{L.~M.} \bibnamefont{Widrow}},
   \bibinfo{journal}{Phys. Rev. Lett.} \textbf{\bibinfo{volume}{72}},
   \bibinfo{pages}{17} (\bibinfo{year}{1994}).

\bibitem[{\citenamefont{Boyarsky et~al.}(2009)\citenamefont{Boyarsky,
 Ruchayskiy, and Shaposhnikov}}]{Boyarsky:2009ix}
\bibinfo{author}{\bibfnamefont{A.}~\bibnamefont{Boyarsky}},
   \bibinfo{author}{\bibfnamefont{O.}~\bibnamefont{Ruchayskiy}},
 \bibnamefont{and}
   \bibinfo{author}{\bibfnamefont{M.}~\bibnamefont{Shaposhnikov}},
   \bibinfo{journal}{Annu. Rev. Nucl. Part. Sci.} \textbf{\bibinfo{volume}{59}},
   \bibinfo{pages}{191} (\bibinfo{year}{2009}).

\bibitem[{\citenamefont{Drewes et~al.}(2017)}]{Adhikari:2016bei}
\bibinfo{author}{\bibfnamefont{M.}~\bibnamefont{Drewes}} \bibnamefont{et~al.},
   \bibinfo{journal}{JCAP} \textbf{\bibinfo{volume}{01}},   \bibinfo{pages}{025}
 (\bibinfo{year}{2017}).

\bibitem[{\citenamefont{Abazajian}(2017)}]{Abazajian:2017tcc}
\bibinfo{author}{\bibfnamefont{K.~N.} \bibnamefont{Abazajian}},
   \bibinfo{journal}{Phys. Rep.} \textbf{\bibinfo{volume}{711-712}},
   \bibinfo{pages}{1} (\bibinfo{year}{2017}).

\bibitem[{\citenamefont{Bernal et~al.}(2017)\citenamefont{Bernal, Heikinheimo,
 Tenkanen, Tuominen, and Vaskonen}}]{Bernal:2017kxu}
\bibinfo{author}{\bibfnamefont{N.}~\bibnamefont{Bernal}},
   \bibinfo{author}{\bibfnamefont{M.}~\bibnamefont{Heikinheimo}},
   \bibinfo{author}{\bibfnamefont{T.}~\bibnamefont{Tenkanen}},
   \bibinfo{author}{\bibfnamefont{K.}~\bibnamefont{Tuominen}}, \bibnamefont{and}
   \bibinfo{author}{\bibfnamefont{V.}~\bibnamefont{Vaskonen}},
   \bibinfo{journal}{Int. J. Mod. Phys. A} \textbf{\bibinfo{volume}{32}},
   \bibinfo{pages}{1730023} (\bibinfo{year}{2017}).

\bibitem[{\citenamefont{Peccei and Quinn}(1977{\natexlab{a}})}]{Peccei:1977ur}
\bibinfo{author}{\bibfnamefont{R.~D.} \bibnamefont{Peccei}} \bibnamefont{and}
   \bibinfo{author}{\bibfnamefont{H.~R.} \bibnamefont{Quinn}},
   \bibinfo{journal}{Phys. Rev. D} \textbf{\bibinfo{volume}{16}},
   \bibinfo{pages}{1791} (\bibinfo{year}{1977}{\natexlab{a}}).

\bibitem[{\citenamefont{Peccei and Quinn}(1977{\natexlab{b}})}]{Peccei:1977hh}
\bibinfo{author}{\bibfnamefont{R.~D.} \bibnamefont{Peccei}} \bibnamefont{and}
   \bibinfo{author}{\bibfnamefont{H.~R.} \bibnamefont{Quinn}},
   \bibinfo{journal}{Phys. Rev. Lett.} \textbf{\bibinfo{volume}{38}},
   \bibinfo{pages}{1440} (\bibinfo{year}{1977}{\natexlab{b}}).

\bibitem[{\citenamefont{Weinberg}(1978)}]{Weinberg:1977ma}
\bibinfo{author}{\bibfnamefont{S.} \bibnamefont{Weinberg}},
   \bibinfo{journal}{Phys. Rev. Lett.} \textbf{\bibinfo{volume}{40}},
   \bibinfo{pages}{223} (\bibinfo{year}{1978}).
 
\bibitem[{\citenamefont{Wilczek}(1978)}]{Wilczek:1977pj}
\bibinfo{author}{\bibfnamefont{F.} \bibnamefont{Wilczek}},
   \bibinfo{journal}{Phys. Rev. Lett.} \textbf{\bibinfo{volume}{40}},
   \bibinfo{pages}{279} (\bibinfo{year}{1978}).

\bibitem[{\citenamefont{Chikashige et~al.}(1981)\citenamefont{Chikashige,
 Mohapatra, and Peccei}}]{Chikashige:1980ui}
\bibinfo{author}{\bibfnamefont{Y.}~\bibnamefont{Chikashige}},
   \bibinfo{author}{\bibfnamefont{R.~N.} \bibnamefont{Mohapatra}},
 \bibnamefont{and}   \bibinfo{author}{\bibfnamefont{R.~D.}
 \bibnamefont{Peccei}},   \bibinfo{journal}{Phys. Lett. B}
 \textbf{\bibinfo{volume}{98}},   \bibinfo{pages}{265} (\bibinfo{year}{1981}).

\bibitem[{\citenamefont{{Zel'dovich} and
 {Novikov}}(1967)}]{1967SvA....10..602Z}
\bibinfo{author}{\bibfnamefont{Y.~B.} \bibnamefont{{Zel'dovich}}}
 \bibnamefont{and}   \bibinfo{author}{\bibfnamefont{I.~D.}
 \bibnamefont{{Novikov}}},   \bibinfo{journal}{Sov. Astron.}
 \textbf{\bibinfo{volume}{10}},   \bibinfo{pages}{602} (\bibinfo{year}{1967}).

\bibitem[{\citenamefont{Carr and Hawking}(1974)}]{Carr:1974nx}
\bibinfo{author}{\bibfnamefont{B.~J.} \bibnamefont{Carr}} \bibnamefont{and}
   \bibinfo{author}{\bibfnamefont{S.~W.} \bibnamefont{Hawking}},
   \bibinfo{journal}{Mon. Not. R. Astron. Soc.} \textbf{\bibinfo{volume}{168}},
   \bibinfo{pages}{399} (\bibinfo{year}{1974}).

\bibitem[{\citenamefont{Witten}(1984)}]{Witten:1984rs}
\bibinfo{author}{\bibfnamefont{E.}~\bibnamefont{Witten}},
   \bibinfo{journal}{Phys. Rev. D} \textbf{\bibinfo{volume}{30}},
   \bibinfo{pages}{272} (\bibinfo{year}{1984}).

\bibitem[{\citenamefont{Lynn et~al.}(1990)\citenamefont{Lynn, Nelson, and
 Tetradis}}]{Lynn:1989xb}
\bibinfo{author}{\bibfnamefont{B.~W.} \bibnamefont{Lynn}},
   \bibinfo{author}{\bibfnamefont{A.~E.} \bibnamefont{Nelson}},
 \bibnamefont{and}   \bibinfo{author}{\bibfnamefont{N.}~\bibnamefont{Tetradis}},
   \bibinfo{journal}{Nucl. Phys. B} \textbf{\bibinfo{volume}{345}},
   \bibinfo{pages}{186} (\bibinfo{year}{1990}).

\bibitem[{\citenamefont{Ricotti and Gould}(2009)}]{Ricotti:2009bs}
\bibinfo{author}{\bibfnamefont{M.}~\bibnamefont{Ricotti}} \bibnamefont{and}
   \bibinfo{author}{\bibfnamefont{A.}~\bibnamefont{Gould}},
   \bibinfo{journal}{Astrophys. J.} \textbf{\bibinfo{volume}{707}},
   \bibinfo{pages}{979} (\bibinfo{year}{2009}).

\bibitem[{\citenamefont{{Chapline}}(1975)}]{1975Natur.253..251C}
\bibinfo{author}{\bibfnamefont{G.~F.} \bibnamefont{{Chapline}}},
   \bibinfo{journal}{Nature (London)} \textbf{\bibinfo{volume}{253}},
   \bibinfo{pages}{251} (\bibinfo{year}{1975}).

\bibitem[{\citenamefont{Carr et~al.}(2010)\citenamefont{Carr, Kohri, Sendouda,
 and Yokoyama}}]{Carr:2009jm}
\bibinfo{author}{\bibfnamefont{B.~J.} \bibnamefont{Carr}},
   \bibinfo{author}{\bibfnamefont{K.}~\bibnamefont{Kohri}},
   \bibinfo{author}{\bibfnamefont{Y.}~\bibnamefont{Sendouda}}, \bibnamefont{and}
   \bibinfo{author}{\bibfnamefont{J.}~\bibnamefont{Yokoyama}},
   \bibinfo{journal}{Phys. Rev. D} \textbf{\bibinfo{volume}{81}},
   \bibinfo{pages}{104019} (\bibinfo{year}{2010}).

\bibitem[{\citenamefont{Bird et~al.}(2016)\citenamefont{Bird, Cholis,
 Mu{\~n}oz, Ali-Ha{\"\i}moud, Kamionkowski, Kovetz, Raccanelli, and
 Riess}}]{Bird:2016dcv}
\bibinfo{author}{\bibfnamefont{S.}~\bibnamefont{Bird}},
   \bibinfo{author}{\bibfnamefont{I.}~\bibnamefont{Cholis}},
   \bibinfo{author}{\bibfnamefont{J.~B.} \bibnamefont{Mu{\~n}oz}},
   \bibinfo{author}{\bibfnamefont{Y.}~\bibnamefont{Ali-Ha{\"\i}moud}},
   \bibinfo{author}{\bibfnamefont{M.}~\bibnamefont{Kamionkowski}},
   \bibinfo{author}{\bibfnamefont{E.~D.} \bibnamefont{Kovetz}},
   \bibinfo{author}{\bibfnamefont{A.}~\bibnamefont{Raccanelli}},
 \bibnamefont{and}   \bibinfo{author}{\bibfnamefont{A.~G.} \bibnamefont{Riess}},
   \bibinfo{journal}{Phys. Rev. Lett.} \textbf{\bibinfo{volume}{116}},
   \bibinfo{pages}{201301} (\bibinfo{year}{2016}).

\bibitem[{\citenamefont{Carr et~al.}(2016)\citenamefont{Carr, Kuhnel, and
 Sandstad}}]{Carr:2016drx}
\bibinfo{author}{\bibfnamefont{B.}~\bibnamefont{Carr}},
   \bibinfo{author}{\bibfnamefont{F.}~\bibnamefont{Kuhnel}}, \bibnamefont{and}
   \bibinfo{author}{\bibfnamefont{M.}~\bibnamefont{Sandstad}},
   \bibinfo{journal}{Phys. Rev. D} \textbf{\bibinfo{volume}{94}},
   \bibinfo{pages}{083504} (\bibinfo{year}{2016}).

\bibitem[{\citenamefont{Clesse and Garc{\'\i}a-Bellido}(2017)}]{Clesse:2016vqa}
\bibinfo{author}{\bibfnamefont{S.}~\bibnamefont{Clesse}} \bibnamefont{and}
   \bibinfo{author}{\bibfnamefont{J.}~\bibnamefont{Garc{\'\i}a-Bellido}},
   \bibinfo{journal}{Phys. Dark Universe} \textbf{\bibinfo{volume}{15}},
   \bibinfo{pages}{142} (\bibinfo{year}{2017}).

\bibitem[{\citenamefont{Green}(2016)}]{Green:2016xgy}
\bibinfo{author}{\bibfnamefont{A.~M.} \bibnamefont{Green}},
   \bibinfo{journal}{Phys. Rev. D} \textbf{\bibinfo{volume}{94}},
   \bibinfo{pages}{063530} (\bibinfo{year}{2016}).

\bibitem[{\citenamefont{Kuhnel and Freese}(2017)}]{Kuhnel:2017pwq}
\bibinfo{author}{\bibfnamefont{F.}~\bibnamefont{Kuhnel}} \bibnamefont{and}
   \bibinfo{author}{\bibfnamefont{K.}~\bibnamefont{Freese}},
   \bibinfo{journal}{Phys. Rev. D} \textbf{\bibinfo{volume}{95}},
   \bibinfo{pages}{083508} (\bibinfo{year}{2017}).

\bibitem[{\citenamefont{Carr et~al.}(2017{\natexlab{a}})\citenamefont{Carr,
 Raidal, Tenkanen, Vaskonen, and Veerm{\"a}e}}]{Carr:2017jsz}
\bibinfo{author}{\bibfnamefont{B.}~\bibnamefont{Carr}},
   \bibinfo{author}{\bibfnamefont{M.}~\bibnamefont{Raidal}},
   \bibinfo{author}{\bibfnamefont{T.}~\bibnamefont{Tenkanen}},
   \bibinfo{author}{\bibfnamefont{V.}~\bibnamefont{Vaskonen}}, \bibnamefont{and}
   \bibinfo{author}{\bibfnamefont{H.}~\bibnamefont{Veerm{\"a}e}},
   \bibinfo{journal}{Phys. Rev. D} \textbf{\bibinfo{volume}{96}},
   \bibinfo{pages}{023514} (\bibinfo{year}{2017}{\natexlab{a}}).

\bibitem[{\citenamefont{Carr et~al.}(2017{\natexlab{b}})\citenamefont{Carr,
 Tenkanen, and Vaskonen}}]{Carr:2017edp}
\bibinfo{author}{\bibfnamefont{B.}~\bibnamefont{Carr}},
   \bibinfo{author}{\bibfnamefont{T.}~\bibnamefont{Tenkanen}}, \bibnamefont{and}
   \bibinfo{author}{\bibfnamefont{V.}~\bibnamefont{Vaskonen}},
   \bibinfo{journal}{Phys. Rev. D} \textbf{\bibinfo{volume}{96}},
   \bibinfo{pages}{063507} (\bibinfo{year}{2017}{\natexlab{b}}).

\bibitem[{\citenamefont{Carr and Kuhnel}(2019)}]{Carr:2018poi}
\bibinfo{author}{\bibfnamefont{B.}~\bibnamefont{Carr}} \bibnamefont{and}
   \bibinfo{author}{\bibfnamefont{F.}~\bibnamefont{Kuhnel}},
   \bibinfo{journal}{Phys. Rev. D} \textbf{\bibinfo{volume}{99}},
   \bibinfo{pages}{103535} (\bibinfo{year}{2019}).

\bibitem[{\citenamefont{Kuhnel and Freese}(2019)}]{Kuhnel:2019xes}
\bibinfo{author}{\bibfnamefont{F.}~\bibnamefont{Kuhnel}} \bibnamefont{and}
   \bibinfo{author}{\bibfnamefont{K.}~\bibnamefont{Freese}},
 arXiv:1906.02744 [gr-qc].
 
\bibitem[{\citenamefont{Carr et~al.}(2019)\citenamefont{Carr, Clesse, Garcia-Bellido, and Kuhnel}}]{Carr:2019kxo}
\bibinfo{author}{\bibfnamefont{B.}~\bibnamefont{Carr}},
   \bibinfo{author}{\bibfnamefont{S.}~\bibnamefont{Clesse}},
   \bibinfo{author}{\bibfnamefont{J.}~\bibnamefont{Garcia-Bellido}} \bibnamefont{and}
   \bibinfo{author}{\bibfnamefont{F.}~\bibnamefont{Kuhnel}},
 arXiv:1906.08217 [astro-ph.CO].

\bibitem[{\citenamefont{Abbott et~al.}(2016{\natexlab{a}})}]{Abbott:2016blz}
\bibinfo{author}{\bibfnamefont{B.~P.} \bibnamefont{Abbott}}
 \bibnamefont{et~al.} (\bibinfo{collaboration}{Virgo, LIGO Scientific
 Collaborations}),   \bibinfo{journal}{Phys. Rev. Lett.}
 \textbf{\bibinfo{volume}{116}},   \bibinfo{pages}{061102}
 (\bibinfo{year}{2016}{\natexlab{a}}).

\bibitem[{\citenamefont{Abbott et~al.}(2016{\natexlab{b}})}]{Abbott:2016nmj}
\bibinfo{author}{\bibfnamefont{B.~P.} \bibnamefont{Abbott}}
 \bibnamefont{et~al.} (\bibinfo{collaboration}{Virgo, LIGO Scientific
 Collaborations}),   \bibinfo{journal}{Phys. Rev. Lett.}
 \textbf{\bibinfo{volume}{116}},   \bibinfo{pages}{241103}
 (\bibinfo{year}{2016}{\natexlab{b}}).

\bibitem[{\citenamefont{Bean and Magueijo}(2002)}]{Bean:2002kx}
\bibinfo{author}{\bibfnamefont{R.}~\bibnamefont{Bean}} \bibnamefont{and}
   \bibinfo{author}{\bibfnamefont{J.}~\bibnamefont{Magueijo}},
   \bibinfo{journal}{Phys. Rev. D} \textbf{\bibinfo{volume}{66}},
   \bibinfo{pages}{063505} (\bibinfo{year}{2002}).

\bibitem[{\citenamefont{Kuhnel and Ohlsson}(2017)}]{Kuhnel:2017ofn}
\bibinfo{author}{\bibfnamefont{F.}~\bibnamefont{Kuhnel}} \bibnamefont{and}
   \bibinfo{author}{\bibfnamefont{T.}~\bibnamefont{Ohlsson}},
   \bibinfo{journal}{Phys. Rev. D} \textbf{\bibinfo{volume}{96}},
   \bibinfo{pages}{103020} (\bibinfo{year}{2017}).

\bibitem[{\citenamefont{Eroshenko}(2016)}]{Eroshenko:2016yve}
\bibinfo{author}{\bibfnamefont{{\relax Yu}.~N.} \bibnamefont{Eroshenko}},
   \bibinfo{journal}{Astron. Lett.} \textbf{\bibinfo{volume}{42}},
   \bibinfo{pages}{347} (\bibinfo{year}{2016}),   \bibinfo{note}{[Pisma Astron.
 Zh. {\bf 42}, 359 (2016)]}.

\bibitem[{\citenamefont{Boucenna et~al.}(2018)\citenamefont{Boucenna, Kuhnel,
 Ohlsson, and Visinelli}}]{Boucenna:2017ghj}
\bibinfo{author}{\bibfnamefont{S.~M.} \bibnamefont{Boucenna}},
   \bibinfo{author}{\bibfnamefont{F.}~\bibnamefont{Kuhnel}},
   \bibinfo{author}{\bibfnamefont{T.}~\bibnamefont{Ohlsson}}, \bibnamefont{and}
   \bibinfo{author}{\bibfnamefont{L.}~\bibnamefont{Visinelli}},
   \bibinfo{journal}{JCAP} \textbf{\bibinfo{volume}{07}},   \bibinfo{pages}{003}
 (\bibinfo{year}{2018}).

\bibitem[{\citenamefont{Adamek et~al.}(2019)}]{Adamek:2019gns}
\bibinfo{author}{\bibfnamefont{J.} \bibnamefont{Adamek}},
\bibinfo{author}{\bibfnamefont{C.~T.} \bibnamefont{Byrnes}},
\bibinfo{author}{\bibfnamefont{M.} \bibnamefont{Gosenca}}, \bibnamefont{and}
\bibinfo{author}{\bibfnamefont{S.} \bibnamefont{Hotchkiss}},
   \bibinfo{journal}{Phys. Rev. D} \textbf{\bibinfo{volume}{100}},
   \bibinfo{pages}{023506} (\bibinfo{year}{2019}).
 
\bibitem[{\citenamefont{Tanabashi et~al.}(2018)}]{Tanabashi:2018oca}
\bibinfo{author}{\bibfnamefont{M.}~\bibnamefont{Tanabashi}}
 \bibnamefont{et~al.} (\bibinfo{collaboration}{Particle Data Group}),
   \bibinfo{journal}{Phys. Rev. D} \textbf{\bibinfo{volume}{98}},
   \bibinfo{pages}{030001} (\bibinfo{year}{2018}).

\bibitem[{\citenamefont{Arias}(2012)}]{Arias:2012az} 
\bibinfo{author}{\bibfnamefont{P.}~\bibnamefont{Arias}},
   \bibinfo{author}{\bibfnamefont{D.}~\bibnamefont{Cadamuro}},
   \bibinfo{author}{\bibfnamefont{M.}~\bibnamefont{Goodsell}},
   \bibinfo{author}{\bibfnamefont{J.}~\bibnamefont{Jaeckel}},
   \bibinfo{author}{\bibfnamefont{J.}~\bibnamefont{Redondo}},
 \bibnamefont{and}   \bibinfo{author}{\bibfnamefont{A.}~\bibnamefont{Ringwald}},
   \bibinfo{journal}{JCAP} \textbf{\bibinfo{volume}{06}},   \bibinfo{pages}{013} (\bibinfo{year}{2012}).

\bibitem[{\citenamefont{Allsman et~al.}(2000)}]{Allsman:2000kg} 
\bibinfo{author}{\bibfnamefont{R.~A.}~\bibnamefont{Allsman}}
 \bibnamefont{et~al.} (\bibinfo{collaboration}{Macho Collaboration}),
   \bibinfo{journal}{Astrophys. J.} \textbf{\bibinfo{volume}{550}},
   \bibinfo{pages}{L169} (\bibinfo{year}{2001}).

\bibitem[{\citenamefont{Tisserand et~al.}(2006)}]{Tisserand:2006zx}
\bibinfo{author}{\bibfnamefont{P.}~\bibnamefont{Tisserand}}
 \bibnamefont{et~al.} (\bibinfo{collaboration}{EROS-2 Collaboration}),
   \bibinfo{journal}{Astron. Astrophys.} \textbf{\bibinfo{volume}{469}},
   \bibinfo{pages}{387} (\bibinfo{year}{2007}).

\bibitem[{\citenamefont{Wyrzykowski et~al.}(2011)}]{Wyrzykowski:2011}
\bibinfo{author}{\bibfnamefont{L.}~\bibnamefont{Wyrzykowski}}
 \bibnamefont{et~al.} (\bibinfo{collaboration}{OGLE Collaboration}),
   \bibinfo{journal}{MNRAS} \textbf{\bibinfo{volume}{416}},
   \bibinfo{pages}{2949} (\bibinfo{year}{2011}).

\bibitem[{\citenamefont{Inoue}(2017)}]{Inoue:2017csr}
\bibinfo{author}{\bibfnamefont{Y.}~\bibnamefont{Inoue}} \bibnamefont{and}
   \bibinfo{author}{\bibfnamefont{A.}~\bibnamefont{Kusenko}},
   \bibinfo{journal}{JCAP} \textbf{\bibinfo{volume}{10}},
   \bibinfo{pages}{034} (\bibinfo{year}{2017}).

\end{thebibliography}
\end{document}